\documentclass[11pt]{article} 
\usepackage{hyperref} 
\usepackage{graphicx}

\begin{document} 
 
\title{Freezing Splashes} 
 
\author{G. Delon, D. Terwagne, N. Adami, A. Bronfort, N. Vandewalle, \\ S. Dorbolo and H. Caps
\\\vspace{6pt} GRASP\\
Physics Department B5 \\
University of Li\`ege \\
B-4000 Li\`ege \\
Belgium } 
 
\maketitle

\begin{abstract} 
We have studied the splashing dynamics of water drops impacting granular layers.
Depending on the drop kinetic energy, various shapes are observed for
the resulting craters. Experimental parameters that have been
considered are : the size of the millimetric droplets ;  the
height of the free fall, ranging from $1.5$~cm to $100$~cm ; and
the diameter of the grains.
As the drop is impacting the granular layer, energy is dissipated and
a splash of grain occurs. Meanwhile, surface tension, inertia and
viscosity compete, leading to strong deformations of the drop
which depend on the experimental conditions. Just after the drop
enters into contact with the granular bed, imbibition takes place and
increases the apparent viscosity of the fluid. The drop
motion is stopped by this phenomenon.
Images and fast-video recordings of the impacts allowed to find
scaling laws for the crater morphology and size.
 \\
This abstract is related to a fluid dynamics video for the APS DFD gallery of fluid motion 2010.
\end{abstract} 
 

It is rather interesting to read the article by A.M. Worthington  \cite{worthington}. The subject is the splash of droplets. The paper is one of the first work that concerns fast imaging and analysis of the impact of a droplet on a solid surface or a pool of liquid. Since this article, the impact of a droplet on a pool of liquid has generated a large number of publications, particularly about the corona formation \cite{brunet} and the bell formed by the impact of large droplet \cite{thorodsen}.
The splash of a droplet on a solid surface is also an active topic of research, which finds applications in numerous industrial areas like painting, self-cleaning windows, impact on airplane wings, agriculture pulverization... From the fundamental science point of view, the splash of a droplet on a solid surface has been revisited thanks to the fine control of the surface texturation and chemical nature. Custom-made Fakir surfaces have been intensively investigated since the splash can be really controlled \cite{callies}. Finally, it is also remarkable that the impact of a solid sphere on a liquid pool has also been recently revisited. Studies on the impact of a solid sphere on a fluidized granular bed also have been reported ({\it e.g.} \cite{deboeuf}), emphasizing the ejection of grains and the formation of craters. The complexity of theses experiences lies in the fact that granular materials are known to exhibit behaviors similar to solids, liquids or gas depending on the constraints. 

\

\begin{figure}
\begin{center}
\includegraphics[width=12.5cm]{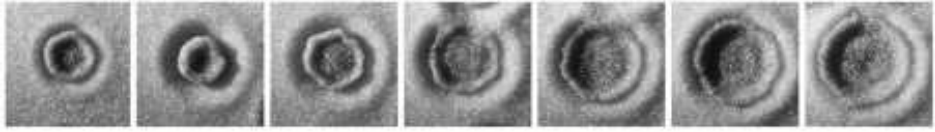}
\caption{Different craters given by droplets impacting a granular layer from different initial heights : between each picture, from the left to the right, the height of dropping is increased by $12$~cm}
\label{impacts}
\end{center}
\end{figure}

In our present study, an apparently common situation is approached: the impact of rain droplets on the beach. The problem is defined by a large number of parameters: droplet size, viscosity, surface tension, impact speed, grain size, compacity. To simplify the problem, we have chosen to vary only the impact speed of the drop and we especially focused on the shape of the crater obtained after the impact and the dynamic of this splash (see samples in Fig.~\ref{impacts}).

The rich morphology of the craters results from the competition between capillary forces, inertia and gravitational drainage. The final crater results from a first phase of spreading, mainly governed by the droplet kinetic energy, before a second stage of retraction, which is driven by surface tension. The dynamics of retraction is affected by the increasing viscosity coming from the melting of sand and water: the viscosity becomes so important that the splash is observed to ``freeze'' during the retraction. Estimation of viscosity remains a challenge and would help in deeply explain the dynamics of this so particular splash.

The first part of the proposed videos present ``classical'' impacts: liquids impacting solids or liquids and solid spheres impacting granular media. The second part presents three representative regimes of frozen splashes : the `donut', the `pie' and the `pancake'. These names are linked to the final shape of the aggregates.  The first one is the result of a droplet impacting the sand with a small velocity. The aggregate has a torus-like shape. The second one, for a medium impacting speed, is flat with a circular rim. This is caled the pie. The third pattern comes from a high velocity impact and is totally flat: this is the pancake. We conclude with an extreme case where the grain size (chocolate powder) is very small and presents a super-hydrophobic effect for the droplet of milk wich is finally mixing with the chocolate powder.

SD and NA thank F.R.S.-F.N.R.S. for financial support. GD benefits a Prodex (Belspo) Grant.

%

\begin{thebibliography}{99}
\bibitem{worthington} A.M.~Worthington, A Study of Splashes, (1908, London, Longmans, Green).
\bibitem{brunet}R.D.~Deegan, P.~Brunet and J.~Eggers, {\em Nonlinearity} {\bf 21},  C1 (2008).
\bibitem{thorodsen} S.T.~Thoroddsen, {\em  J. Fluid Mech} {\bf 451}, 373 (2002). 
\bibitem{callies} M.~Callies and  D.Qu\'er\'e, {\em Soft Matter} {\bf 1}, 55 (2005) .
\bibitem{deboeuf} S.~Deboeuf, P.~Gondret and M.~Rabaud, {\em Phys. Rev. E} {\bf 79}, 041306 (2009).

 \end{thebibliography}
\end{document}